\documentclass[%
 twocolumn,
 aps, prapplied, longbibliography,
 amsmath,amssymb,
 superscriptaddress
]{revtex4-1}

\usepackage{graphicx}
\usepackage{dcolumn}
\usepackage{bm}

\usepackage[utf8]{inputenc}
\usepackage[T1]{fontenc}
\usepackage{mathptmx}
\usepackage{graphicx}
\usepackage{dcolumn}
\usepackage{bm}
\usepackage{xcolor}
\definecolor{blue}{rgb}{0,0.5,1.0}
\usepackage[export]{adjustbox}
\usepackage{braket}
\usepackage{amsmath}
\usepackage{MnSymbol}

\begin{document}

\title[Proposal of a realistic stochastic rotor engine based on electron shuttling]{Proposal of a realistic stochastic rotor engine based on electron shuttling}

\author{C. W. W\"achtler}
\email{christopher.w.waechtler@campus.tu-berlin.de}
\affiliation{Institute of Theoretical Physics, Secr. EW 7-1, Technical University Berlin, Hardenbergstr. 36, D-10623 Berlin, Germany}
\author{P. Strasberg}
\affiliation{F\' isica Te\` orica: Informaci\' o i Fen\` omens Qu\` antics, Departament de F\' isica, Universitat Aut\` onoma de Barcelona, ES-08193 Bellaterra (Barcelona), Spain}
\author{G. Schaller}
\affiliation{Institute of Theoretical Physics, Secr. EW 7-1, Technical University Berlin, Hardenbergstr. 36, D-10623 Berlin, Germany}

\date{\today}

\begin{abstract}
 We propose and analyse an autonomous engine, which combines ideas from electronic transport and self-oscillating 
 heat engines. It is based on the electron-shuttling mechanism in conjunction with a rotational degree of freedom. 
 We focus in particular on the isothermal regime, where chemical work is converted into mechanical work or vice versa, 
 and we especially pay attention to use parameters estimated from experimental data of available single components. Our analysis shows that for 
 these parameters the engine already works remarkably stable, albeit with moderate efficiency. Moreover, it has the advantage that it can be up-scaled to increase power and reduce fluctuations further.
\end{abstract}

\maketitle

\section{\label{sec:Introduction}Introduction}
A major challenge in a world with rapidly growing nanotechnological abilities is to understand and design 
efficient and realizable micro-machines in form of heat pumps, refrigerators or isothermal engines. Particularly 
promising examples are so-called \emph{autonomous} machines, which do not require any active regulation from the 
outside. Here, two different approaches seem to be outstanding. First, thermoelectric devices, which use the interplay 
of thermal and chemical gradients to perform useful tasks, were proposed \cite{SegalPRL2008, EspositoEtAlEPL2009, SanchezButtikerPRB2011, StrasbergEtAlPRL2013, SothmannEtAlNano2014, BenentiEtAlPhysRep2017} and experimentally realized using quantum dot (QD) structures \cite{FeshchenkoEtAlPRB2014, HartmannEtAlPRL2015, ThierschmannEtAllNano2015, JosefssonEtAlNature2018}. Second, self-oscillating machines, which are coupled to multiple thermal reservoirs, were analyzed 
theoretically \cite{FilligerReimannPRL2007, WangVukoivPRL2008, SmirnovEtAlNanotech2009, AlickiEtAlJPA2015, AlickiJPA2016, AlickiEntropy2016, AlickiEtAlAP2017, RouletEtAlPRE2017, CerasoliEtAlPRE2018, SeahEtAlNJP2018} and experimentally \cite{ChianEtAlPRE2017, SoniEtAl2017, SerraGarciaEtAlPRL2016}. While moving parts can be challenging to implement in nanoscale systems, self-oscillating machines offer the possibility to study the use and conversion of mechanical work within an autonomous setting that does not rely on time-dependent control fields.\\

Here, we propose an electrostatic DC (direct current) engine based on single electron tunneling, which combines ideas from both areas. Its design is close to the 
conventional electron shuttle, which is well studied in theory \cite{GorelikEtAlPRL1998, IsacssonEtAlPhysicaB1998, FedoretsEtAlEPL2002, NovotnyEtAlPRL2003, NovotnyEtAlPRL2004, UtamiEtAlPRB2006, NoceraEtAlPRB2011} and practice \cite{ParkEtAlNature2000, ScheibleBlickAPL2004, MoskalenkoEtAlPRB2009, MoskalenkoEtAlNanotechnology2009, KonigWeigAPL2012} and which was recently also investigated from a thermodynamic perspective \cite{TonekaboniEtAlArXiv2018, WaechtlerEtAl}. However, instead of using a harmonic oscillator to shuttle electrons between two reservoirs, we will use a rotational degree of freedom \cite{CroyEisfeldEPL2012, CelestinoEtAlNJP2016}. To have an unambiguous notion of mechanical work, a liftable weight is attached to this rotational degree of freedom. We provide a thorough theoretical and numerical analysis of the power output and efficiency in the isothermal regime, where chemical work is converted into mechanical work (as in a car) or vice versa (as in a turbine). Another main point is to pay attention to use experimentally realistic parameters. Thus, we demonstrate that our device is implementable with state-of-the-art technologies. Furthermore, it is possible to scale up our engine by using multiple quantum dots attached at appropriate positions to the same rotational degree of freedom. 

\section{\label{sec:Model}Model}
\begin{figure}
\includegraphics[width=\columnwidth]{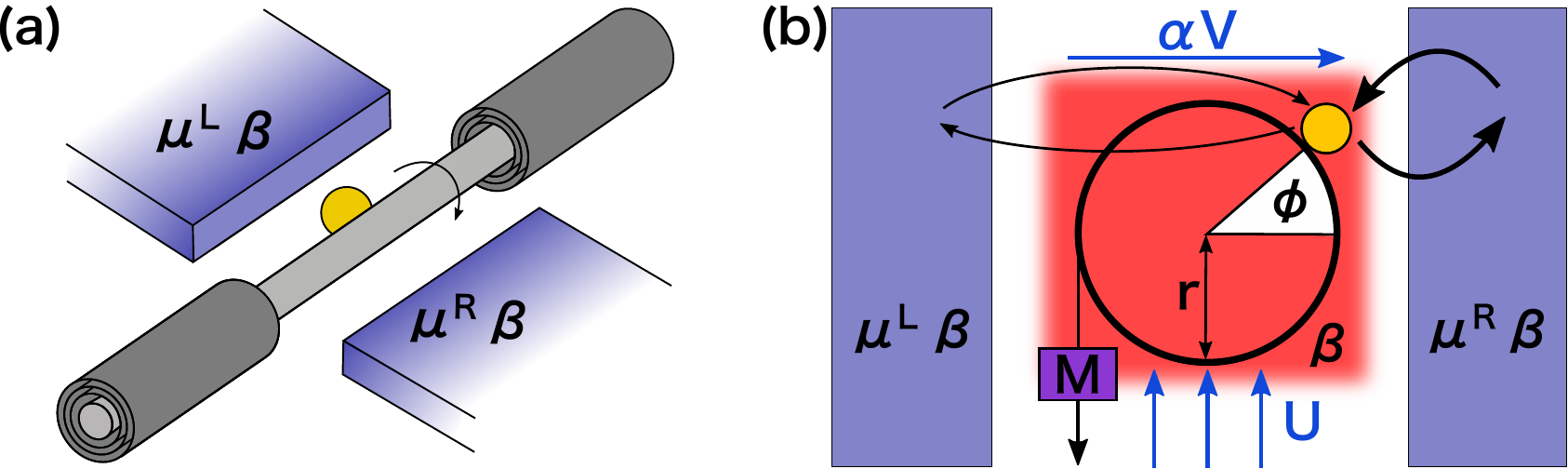}
\caption{(a) Proposed realization with a gold particle mounted onto a carbon nanotube (see main text). (b) Physically relevant quantities of the engine (details in the main text).}
\label{fig:WheelEngine}
\end{figure}

The model is depicted in Fig.~\ref{fig:WheelEngine}. As the rotational degree of freedom (called `rotor' in the 
following) we imagine a multi-walled carbon nanotube, where the outer walls have been removed using electrical 
breakdown techniques~\cite{CollinsEtAlScience2001, CollinsEtAlPRL2001}. 
Then, the inner shell with radius $r$ and moment of inertia $I$ can be accessed and rotate while being held by 
the outer walls. Interestingly, such a bearing has been realized experimentally \cite{FennimoreEtAlNature2003, BourlonEtAlNano2004}. We suggest to mount a gold nanoparticle onto the nanotube, e.g. using dip-pen nanolithography \cite{ChuEtAlJPCC2008, ChuEtAlCCR2010, MartyEtAl2006}, which serves as a QD with on-site energy $\varepsilon$. Similar to the electron shuttle, the QD is tunnel-coupled to two leads with chemical potentials $\mu^{\text{L}} = \varepsilon + \text{e}V/2$ and $\mu^{\text{R}} = \varepsilon - \text{e}V/2$ for the left and right lead, respectively, at inverse temperature $\beta$. Here, $V$ denotes the applied bias voltage between the two leads. We idealize the QD to a single level (Coulomb blockade \cite{GiaeverZellerPRL1968, KulikShekhter1975, AverinLikharevJLTP1986}), such that the QD is either empty ($q=0$) or occupied by exactly one electron ($q=\text{e}$). It is to be expected, however, that lifting the assumption of Coulomb blockade does not decrease the thermodynamic performance of the engine, compare also with Sec.~\ref{sec:Upscaling}. The movement of the rotor is described in one dimension with angle $\phi\in \mathbb R$ and angular velocity $\omega\in \mathbb R$, where $\phi=0$ is the rightmost position of the QD and increases with anti-clockwise rotation [see Fig.~\ref{fig:WheelEngine} (b)]. 

With only the lead bias applied, clockwise and counter-clockwise rotation would be equally likely. It is therefore necessary to break also the top-bottom symmetry. While this symmetry breaking can be achieved with energy-dependent tunneling rates \cite{SegalPRL2008, SanchezButtikerPRB2011, StrasbergEtAlPRL2013} or geometric design \cite{CelestinoEtAlNJP2016}, we aim for a simple approach that does not require any fine-tuning of the device. Therefore, we consider an additional transverse field, such that the coupling of the rotor motion and the charge state is introduced by two perpendicular electric fields $\alpha V$ and $U$. The first is generated by the bias voltage and is assumed to be homogeneous between the  leads~\cite{GorelikEtAlPRL1998}. The second is applied externally and breaks the top-bottom symmetry. Then, an electrostatic torque $\tau_{\text{el}}(\phi,q)=-\alpha V q r \sin(\phi) + U q r \cos(\phi)$ acts on the QD (see Appendix~\ref{secApp:DerivationTorque}). To demonstrate power output, we connect a weight with mass $M$ to the rotor [see Fig.~\ref{fig:WheelEngine} (b)] such that a constant gravitational torque of $\tau_{\text{M}}=M \text gr$ is acting on the rotor. 

The dynamics of the rotor is modeled as underdamped motion with friction constant $\gamma$ such that the rotor may perform a whole revolution due to inertia and stable rotations are possible. The friction arises e.g. from the interaction between the walls of the multi-walled carbon nanotube and the displacement of the image charge in the leads. 
Additionally, the rotor is small, such that thermal fluctuations from a heat bath at inverse temperature $\beta$ have to be taken into account. Assuming negligible gold particle mass compared to the rotor mass, the dynamics of the engine can be described by a coupled Fokker-Planck and master equation, known e.g. from switching diffusion processes \cite{ReinmannPhysRep2002, YangGePRE2018}:
\begin{equation}
\label{eq:FPE}
\begin{aligned}
\frac{\partial p_q}{\partial t} =& \left[-\omega \frac{\partial }{\partial \phi} + \frac{1}{I}\frac{\partial}{\partial\omega}\left(\gamma \omega+DI\frac{\partial}{\partial \omega}\right)\right]p_q\\
&-\frac{1}{I}\frac{\partial}{\partial \omega}\left[\tau_{\text{el}}(\phi,q) + \tau_{\text{M}}\right] p_q
+\sum\limits_{q'\nu}R_{qq'}^\nu(\phi)p_{q'}.
\end{aligned}
\end{equation}
Here, $p_q\equiv p_q(\phi,\omega;t)$ is the joint probability density to find the engine at the state $(\phi,\omega,q)$ at time $t$, and we have introduced a velocity diffusion coefficient $D = \gamma / (\beta I^2)$. The first line of Eq.~(\ref{eq:FPE}) describes the free rotational diffusion of the rotor and the first term in the second line represents the two torque contributions. The rate matrix $R_{qq'}^\nu(\phi)$ describes transitions from state $q'$ to state $q$, i.e., electron tunneling between the QD and the lead $\nu=\{\text{L},\text{R}\}$. The elements of the matrix are fixed by the rates 
\begin{equation}
\label{eq:Rates}
\begin{aligned}
R^{\nu}_{\text{e}0}(\phi)&= \Gamma \exp\left[\pm r\cos(\phi)/\lambda\right]f^{\nu}(\phi),\\
R^{\nu}_{0\text{e}}(\phi)&= \Gamma \exp\left[\pm r\cos(\phi)/\lambda\right]\left[1-f^{\nu}(\phi)\right]
\end{aligned}
\end{equation}
and the probability conservation condition, $R_{qq}^\nu(\phi) = -\sum_{q'\neq q}R_{q'q}^\nu(\phi)$. Here, $\Gamma$ denotes a bare transition rate controlling the overall tunneling time-scale. Since quantum mechanical tunneling is exponentially sensitive to the tunneling distance, the rates are modulated  differently, the $+$ and $-$ signs hold for $\nu=R$ and $\nu=L$, respectively, with characteristic tunneling length 
$\lambda$~\cite{GorelikEtAlPRL1998, LaiEtAlJoP2012, LaiEtAlJoP2013, FedoretsEtAlPRL2004, FedoretsEtAlEPL2002}. Additionally, the rates depend on the probability of an electron (hole) with matching energy in the reservoir, i.e., on the Fermi distribution 
\begin{equation}
f^\nu(\phi) \equiv \frac{1}{\exp\left\{\beta^\nu(\varepsilon-\alpha \text{e}V r \cos(\phi) - \text{e}Ur \sin(\phi)-\mu^\nu)\right\}+1}.
\end{equation} 
Here, the difference in energy between a filled and empty QD in the electrostatic field enters the Fermi function, see e.g. Refs.~\cite{LaiEtAlFoP2015, WaechtlerEtAl} and Appendix~\ref{secApp:Thermodynamics}. 

\section{\label{sec:Parameters}Experimental parameters}
In order for the proposed engine to be realized it is important that the parameters of the theoretical system lie within an experimentally accessible range. Clearly, the engine can also be fabricated in different ways than proposed in this work, however, for the proposed experimental realization there exists data in the literature for estimating the parameters. 

We assume a radius $r=4$nm \cite{KobayashiEtAlCarbon2011} and a length $l=0.6\mu$m \cite{RenEtAllScience1998} of the carbon nanotube, deposited between two electronic leads of distance $d=10$nm, which can be achieved by junction-breaking technique \cite{ParkEtALAPL1999}. Since $\alpha \approx 1/d$, we approximate 
$\alpha = 0.1 \text{nm}^{-1}$. The moment of inertia $I$ of the carbon nanotube with the mentioned dimensions can be approximated to be $I=19.2\cdot 10^{-38} \text{kg m}^2$\cite{LaurentEtAlCarbon2010}. Usual temperatures at which single electron experiments are performed range from mK to a few K \cite{ParkEtAlNature2000, MartyEtAl2006} and we use a temperature of $T=10$K. The applied bias voltage $V$ is given in order of mV \cite{ParkEtAlNature2000}, such that the dimensionless quantity $\beta V$ lies in the order of magnitude of $10$.

Next we estimate the timescale of tunneling. We approximate the rate of tunneling by looking at the tunneling current between gold particles and an STM tip \cite{XuChenCPL2009}. Since the measured current is in the low nA regime, which is roughly $10^9$ electrons per second, the experimental tunneling rate can be estimated with about $\Gamma\approx 10^9/\text{s}$. Then, $\left<\omega\right>/\Gamma$ is of the order of magnitude of $1$. For the characteristic tunneling length we choose a value of $\lambda=3$nm, such that the quotient $r/\lambda$ is in the same order of magnitude as for an electron shuttle consisting of a gold nanoparticle in between two gold electrodes \cite{MoskalenkoEtAlPRB2009}. We assume that the externally applied electric field can be easily adjusted and we choose a value of $U = 4 \text{mV}/$nm. Lastly, we choose $\gamma = 0.8\cdot 10^{-30} \text{kg m}^2/\text{s}$, where the friction $\gamma$ is two orders of magnitude larger than estimated for interaction between the walls of the multi-walled carbon nanotube \cite{ServantieGaspardPRL2006} in order to take additional effects like coupling of the charge on the QD with the image charge on the leads or friction due to the mounted gold particle into account. For a proper functioning as a useful thermodynamic device, the friction constant $\gamma$ and tunneling length $\lambda$ need careful fine-tuning in the experiment. Our simulations suggest that the influence of other parameters is of minor relevance instead.

In the following we vary the applied bias voltage $V$ as well as the mass $M$ attached to the rotor. We imagine 
that these two quantities are the easiest ones to vary in a given experimental setup. 

\section{\label{sec:Dynamics}Dynamics of the rotor engine}

\begin{figure}
\includegraphics[width=\columnwidth]{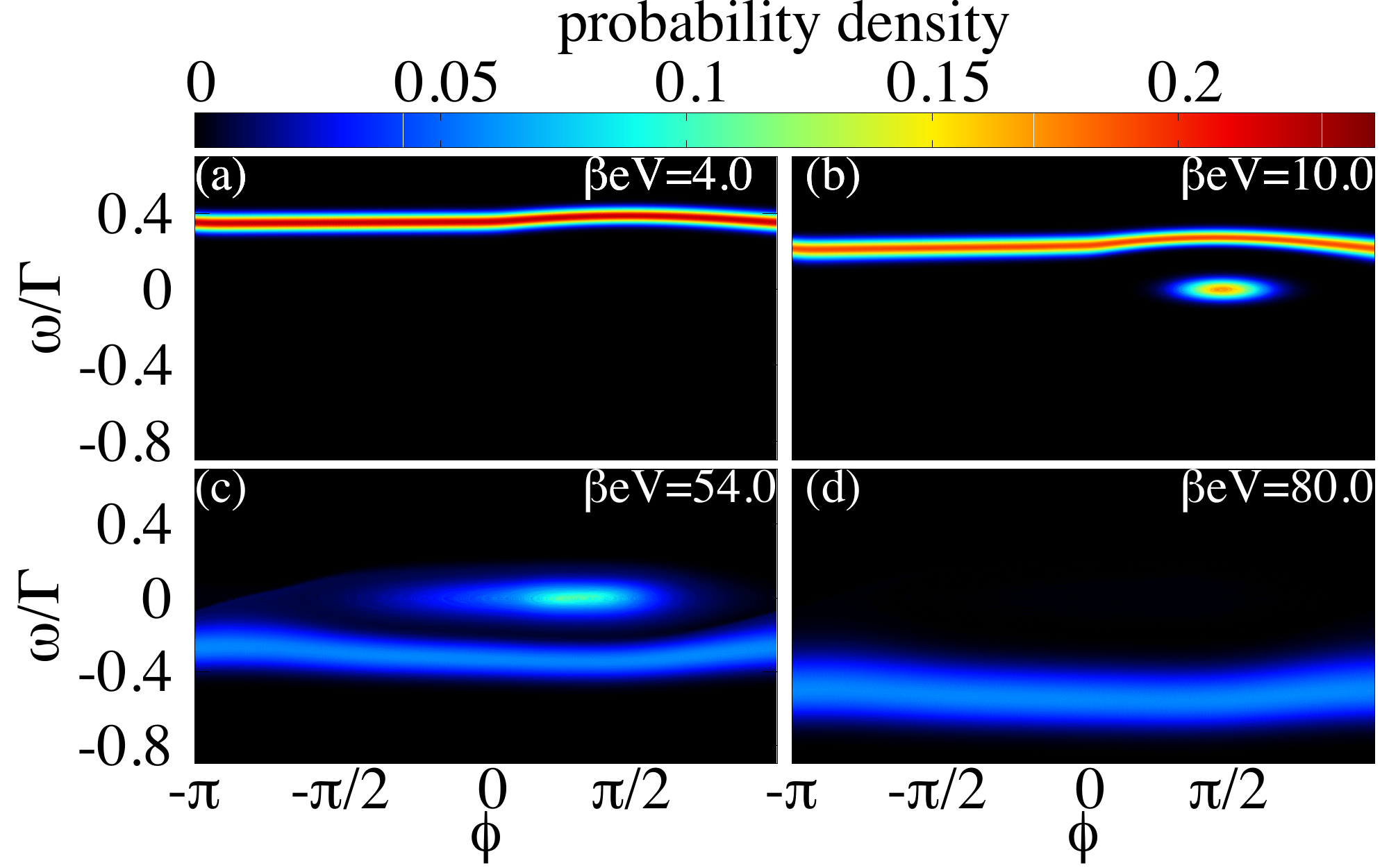}
\caption{Probability density of the rotor $p(\phi,\omega) = \sum_q p(\phi,\omega,q)$ in phase-space (periodic in $\phi$) for $\beta M \text g r = 2.0$ and different values of $\beta \text{e}V$.}
\label{fig:Histogramms}
\end{figure}

To understand the steady state dynamics of the rotor, we show in Fig.~\ref{fig:Histogramms} the stationary probability 
density of the rotor alone [$p(\phi,\omega) = \sum_q p_q(\phi,\omega)$] for different values of $V$ for the case of 
$\beta M \text{g}r = 2.0$. The steady state probability density is obtained by numerically solving the trajectory representation of Eq.~(\ref{eq:FPE}) and assuming periodic boundary conditions of $\phi$ (see Appendix~\ref{secApp:Trajectory} for details). Three basic regimes can be distinguished: 

\textit{Falling weight:}  For a small bias voltage [panel (a) $\beta \text{e}V = 4.0$] the net electric field is almost  pointing along the externally applied electric field ($\beta\text{e}U/\alpha = 40.0$). Thus, the left-right symmetry is almost unbroken and without applied weight, the rotor would not move on average. However, due to the gravitational torque 
it turns counter-clockwise ($\omega >0$). Utilizing this mechanism, the engine can also pump electrons against the bias voltage, which we will discuss later.

\textit{Lifting weight:} A large bias voltage [panel (d) $\beta \text{e}V = 80$] breaks the left-right symmetry  and the rotor turns clockwise ($\omega<0$) with a typical trajectory described as follows: When the QD is close to the left lead, an electron is loaded onto the QD and the electric field exerts a (clockwise) torque on the rotor. As the QD approaches the right lead, the electron is unloaded and due to inertia approaches the left lead again, thereby closing the cycle. Note that lifting the weight relies on stochastic electron jumps at specific moments. Hence, the variance of $\omega$ is increased [Fig.~\ref{fig:Histogramms} (d)] compared to a falling weight [Fig.~\ref{fig:Histogramms} (a)], where $\tau_{\text{M}}$ is exerted independently of $\phi$.

\textit{Bistable regime:} In between the two cases there exists a crossover regime [see Fig.~\ref{fig:Histogramms} (b) and (c)], in which the two operational modes compete. Below a threshold voltage of $\beta \text{e}V_{\text{low}}\approx 11.0$ the rotor turns counter-clockwise due to $M$. However, as $V$ is increased and the symmetry is further broken, $\tau_{\text{el}}$ is competing with $\tau_{\text{M}}$, such that $\left<\omega\right>$ decreases. Simultaneously, a circle around the point $(\phi\approx \pi/2$, $\omega=0$) emerges indicating a standstill of the charged engine ($q=\text{e}$). As $V$ is further increased, solely the rest state survives and above $\beta \text{e}V_{\text{high}}\approx 42.0$, $\tau_{\text{el}}$ overcomes $\tau_{\text{M}}$ and the friction, resulting in the coexistence of standstill and clockwise rotation ($\omega<0$) [see Fig.~\ref{fig:Histogramms} (c)]. Due to the stochasticity of the electron jumps, the circle as well as the line in panel (c) is smeared out compared to panel (b). Notice that, when we would model the dynamics of the rotor at a meanfield level, there would be a sharp transition from a clockwise to a counter-clockwise spinning rotor due to an underlying Hopf bifurcation of the 
nonlinear dynamics \cite{WaechtlerEtAl}. 

\section{\label{sec:Thermodynamics}Work output, efficiency and reliability}
We now turn to the thermodynamic description and focus on the interconversion of the two resources chemical and 
mechanical work at constant temperature. We use the convention that work performed by the system is negative. The average chemical and mechanical power are defined as 
\begin{equation}
\label{eq:DefinitionWork}
\left<\dot W^\text{c}\right> = \sum_\nu \mu^\nu \left<I_{\text{M}}^\nu\right> \quad \text{ and } \quad  \left<\dot W^{\text{m}}\right> = M\text{g}r \left<\omega\right>,
\end{equation}
where  $\left<I_{\text{M}}^\nu\right> = \int d\phi d\omega \left[ R^\nu_{10}(\phi)p_0-R^\nu_{01}(\phi)p_1\right]$ is the matter current coming from lead $\nu$ and $\left<\omega\right>=\sum_q\int d\phi d\omega~\omega p_q(\phi,\omega;t)$. Here, $p_q(\phi,\omega;t)$ is the solution of Eq.~(\ref{eq:FPE}). The second law of thermodynamics at steady state ensures the non-negativity of entropy production rate, 
\begin{equation}
\label{eq:Entropy}
\dot\Sigma = \beta\left(\left<\dot W^\text{c}\right> + \left<\dot W^\text{m}\right>\right) \ge 0,
\end{equation}
which describes the fundamental trade-off between extracting chemical (mechanical) work at the expense of consuming 
mechanical (chemical) work. Its derivation follows from first principles (see Appendix~\ref{secApp:Thermodynamics}). 

\begin{figure}
\includegraphics[width=\columnwidth]{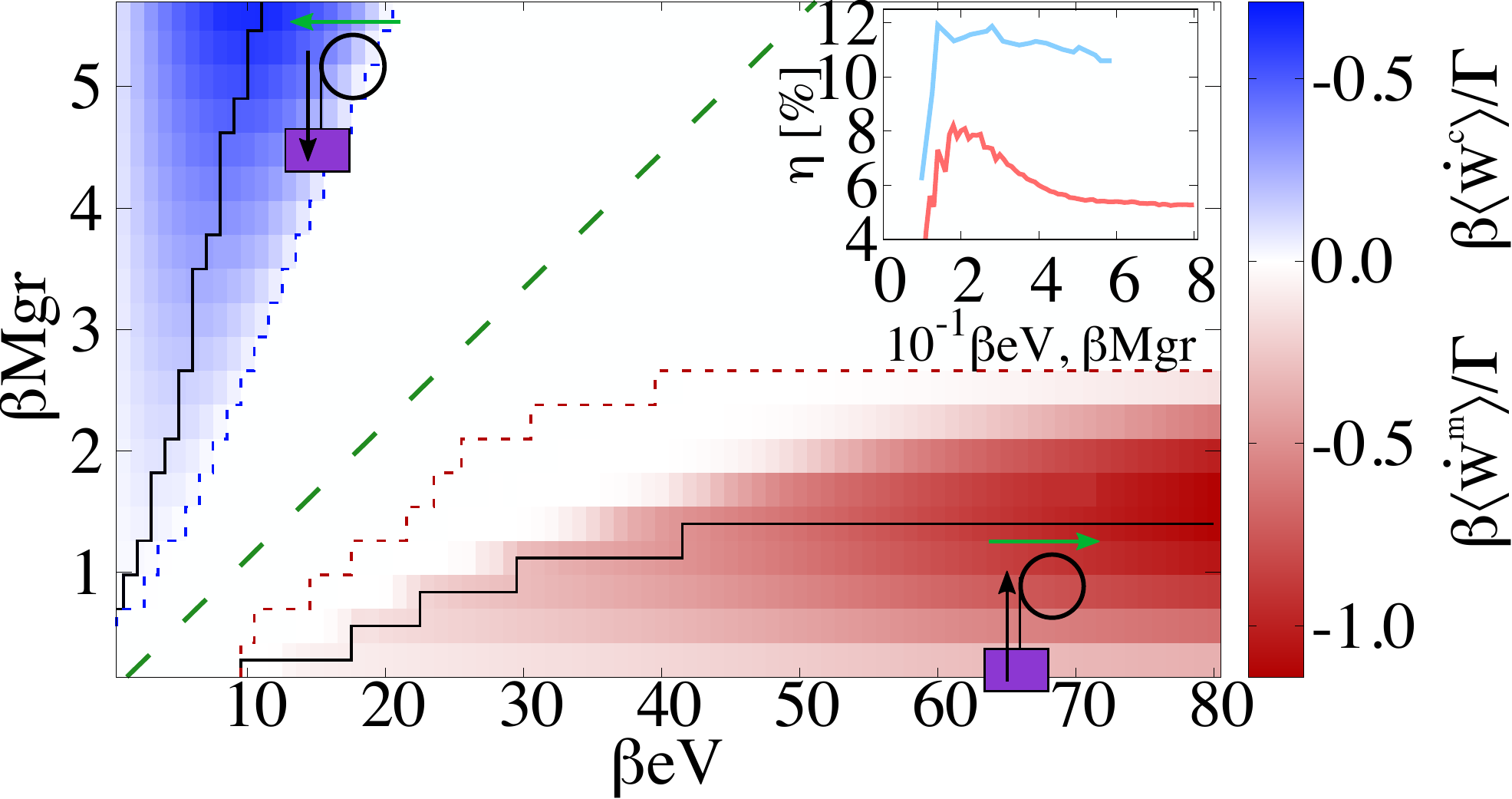}
\caption{Power output for different values of $\beta \text{e}V$ and $\beta M\text{g}r$. Within the region enclosed by the red (blue) dashed line $\left<\dot W^\text m\right> >0$ ($\left<\dot W^\text c\right> >0$). The illustrations indicate the directionality of the weight (black) and the matter current (green). The green dashed line separates the two regimes by thermodynamic arguments (see main text) and the black lines correspond to the maximum power output curves. Inset: Efficiency at maximum power (red/blue for lifting/pumping).}
\label{fig:EnginePumping}
\end{figure}

Fig.~\ref{fig:EnginePumping} demonstrates that our engine works as desired. Within the region enclosed by the red dotted curve of Fig.~\ref{fig:EnginePumping} chemical work is transformed into mechanical work, i.e., the weight is lifted ($\omega <0$). Here, $p(\phi,\omega)$ is similar to the ones shown in Fig.~\ref{fig:Histogramms} (c) and (d), i.e., a coexistence of standstill and clockwise turning or solely clockwise turning. Coexistence, however, is found especially between the red area and the red dotted curve in Fig.~\ref{fig:EnginePumping}, i.e., where $\left<\dot W^{\text{m}}\right> \approx 0$.

On the other hand, the blue area of Fig.~\ref{fig:EnginePumping} corresponds to values of $V$ and $M$ for which electrons are pumped ($\left<\dot W^\text{c}\right><0$) by utilizing mechanical work ($\left<\dot W^{\text{m}}\right> >0$), i.e., the weight is falling ($\omega >0$). In order to understand the electron pumping mechanism, we introduce the effective chemical potential $\tilde \mu^\nu = \mu^\nu+\alpha\text{e}Vr\cos(\phi)+\text{e}Ur\sin(\phi)$, which enters the Fermi functions. For small values of $V$ and large values of $M$, the rotor is turning counter-clockwise ($\omega<0$). For $\phi\in(-\pi/2,\pi/2)$ the QD is more likely to interact with the right lead and $\tilde \mu^{\text{R}}>\varepsilon$. Thus, an electron may tunnel into the QD from the right lead. Consequently, the filled QD approaches the left lead  with $\tilde \mu^{\text{L}}<\varepsilon$ for $\phi\in(\pi/2,3\pi/2)$ and the electron is unloaded into the left lead. Then, the empty dot approaches the right lead again and the cycle is closed, in which one electron has been pumped. As $V$ is increased, $\tilde \mu^{\text{R}}<\varepsilon$ and pumping is no longer possible. 

In the white region in Fig.~\ref{fig:EnginePumping} not enclosed by the red/blue dashed lines, the engine does not perform any work output and the input power is solely dissipated as heat into the different reservoirs. In that case the engine is either at rest ($\beta M \text g r <3$) or turning counter-clockwise ($\beta M\text g r>3$). For infinite bias or infinite mass, the colored regions will vanish (not shown): For $V\to\infty$, the top-down symmetry is no longer broken by the net electric field. For $M\to\infty$, the large rotation velocity leads to an effective single dot picture placed at an average position $\left<r \cos(\phi)\right>=0$, inhibiting electron pumping.

The two regions discussed above are separated not only by the different rotational direction but also by thermodynamical arguments: According to the second law the output power can be at most equal to the (negative) input power. Hence, for electron pumping ($\left<\omega\right>>0$) the bound is given by $\left<\dot W^{\text{c}}\right> \leq -\left<\dot W^{\text{m}}\right>$. Using the fact that at most one electron per revolution can be pumped ($\left<I_{\text{M}}^{\text{L}}\right> \leq - \left<\omega\right>/2\pi$) results in $V \leq 2\pi M\text{g}r$. For $V< 2\pi M\text{g}r$, only pumping is possible and for $V>2\pi M\text{g}r$ only lifting. We plot this bound in Fig.~\ref{fig:EnginePumping} as green dashed line exactly separating the two regimes. 

The performance of the proposed engine can be quantified by the efficiency $\eta = {-\left<\dot W^{\text{out}}\right>}/{\left<\dot W^{\text{in}}\right>}\leq 1$, where the bound follows from Eq.~(\ref{eq:Entropy}). 
Saturation of the bound is only possible for zero power output, i.e. an infinite long cycle. Since our machine operates in finite time, we discuss the \emph{efficiency at maximum power} $\eta_\text{max}$\cite{Chambadal1957, Novikov1958, CurzonAhlbornAJP1975, SchmiedlSeifertEPL2007, SchmiedlSeifertEPL2008, EspositoEtAlEPL2009, EspositoEtAlPRL2010, JosefssonEtAlNature2018}. The inset of Fig.~\ref{fig:EnginePumping} shows this quantity along the two black curves in the main plot, which correspond to the maximum power output within each region. In both cases, the efficiency first increases after which further increase of $V$ (red line) or $M$ (blue line) results in a decreasing efficiency. For experimentally feasible parameters (see Sec.~\ref{sec:Parameters}), $\eta_{\text{max}}^{\text{m}} \approx 8.1\%$ for $\beta \text{e}V = 18.0$ and $\beta M\text{g} r = 0.6$ and $\eta_{\text{max}}^\text{c} \approx 9.4\%$ for $\beta M\text{g}r = 1.4$ and $\beta \text{e}V = 3.0$. 

\begin{figure}
\includegraphics[width=\columnwidth]{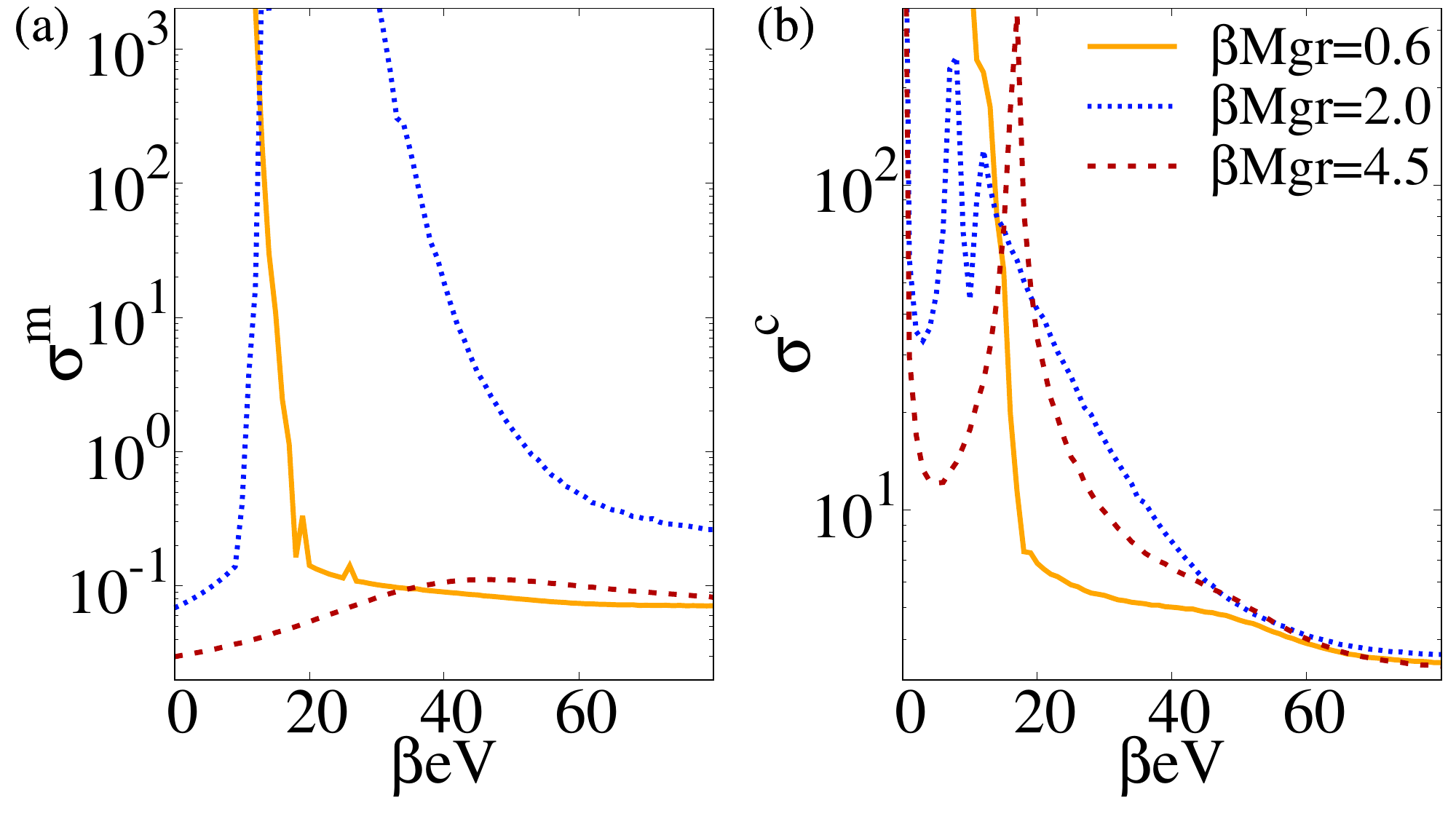}
\caption{Normalized fluctuations $\sigma^{\text{m}}$ (a) and $\sigma^\text{c}$ (b) as function of $\beta \text e V$ for different values of $\beta M \text g r$.}
\label{fig:Noise}
\end{figure}

Apart from efficiency another figure of merit of a stochastic engine is its reliability. This can be described, e.g., 
by the normalized standard deviation
\begin{equation}
\sigma^{\text{m}/\text{c}} \equiv \frac{\sqrt{\left<\left(\dot W^{\text{m}/\text{c}}\right)^2\right>-\left<\dot W^{\text{m}/\text{c}}\right>^2}}{\left|\left<\dot W^{\text{m}/\text{c}}\right>\right|}.
\end{equation}
In Fig.~\ref{fig:Noise} (a) we plot $\sigma^{\text{m}}$ as a function of $V$ for $\beta M \text{g}r = 0.6$ (orange solid), $\beta M\text{g} r = 2.0$ (blue dotted) and $\beta M\text{g}r = 4.5$ (red dashed). For the smallest attached weight (orange solid) there are large fluctuations for $\beta\text{e} V<15.0$. Here, the rotor is at rest, such that $\left<\dot W^{\text{m}}\right> = 0$. Then, $\sigma^{\text{m}}$ diverges.  As the engine is turning clockwise ($\beta \text{e}V >15.0$) the fluctuations decrease. A similar behavior can be observed for $\beta M \text{g}r=2.0$ (blue dotted). Here, below $\beta\text{e} V =12.0$ the fluctuations are small, which corresponds to a falling weight ($\omega>0$). Fluctuations drastically increase as the rotor comes to a standstill and subsequently decrease as the weight is lifted ($\omega<0$). The slow decay of fluctuations results from telegraph noise induced by the switching between lifting and rest state \cite{CottetEtAlPRB2004, JordanSukhorukovPRL2004, SchallerEtAlPRB2009, SchallerEtAlPRB2010}. Along the red dashed line ($\beta M \text{g}r = 4.5$) the rotor is always turning counter-clockwise and, thus, fluctuations do not diverge. However, as electrons are no longer pumped ($\beta \text{e}V>16$) the rotor is decelerated and $\sigma^{\text{m}}$ increases about one order of magnitude.

Fig.~\ref{fig:Noise} (b) shows $\sigma^\text{c}$ for the same values of $M$ as before. For $\beta M\text{g}r = 4.5$ (red dashed) $\sigma^\text{c}$ diverges at $\beta\text{e}V=0.0$ since $\left<\dot W^\text{c}\right> = 0$ [see Eq.~(\ref{eq:DefinitionWork})]. As $V$ is increased, almost in each cycle one electron is pumped as a very regular process and fluctuations decay. Upon further increase of $V$ electron pumping against the bias becomes less likely with decreasing current and as the current switches direction, i.e., electrons follow the descent of the chemical potentials from left to right, $\sigma^\text{c}$ peaks. The latter corresponds to the border of the blue area in Fig.~\ref{fig:EnginePumping}. Further increase of $V$ again results in a regular current (now along the bias) and fluctuations decay. For $\beta M \text{g}r = 0.6$ (orange solid) and small $V$ the rotor is at standstill. As discussed above, the rest state is accompanied by large fluctuations $\sigma^\text{c}$. As the rotor starts to move ($\beta \text{e}V > 15.0$) the current (along the chemical potentials) becomes more regular and fluctuations decrease. For the intermediate weight ($\beta M\text{g} r = 2.0$, blue dotted) all effects discussed so far can be observed. 

\section{\label{sec:Upscaling}Possible upscaling}

\begin{figure}
\includegraphics[width=\columnwidth]{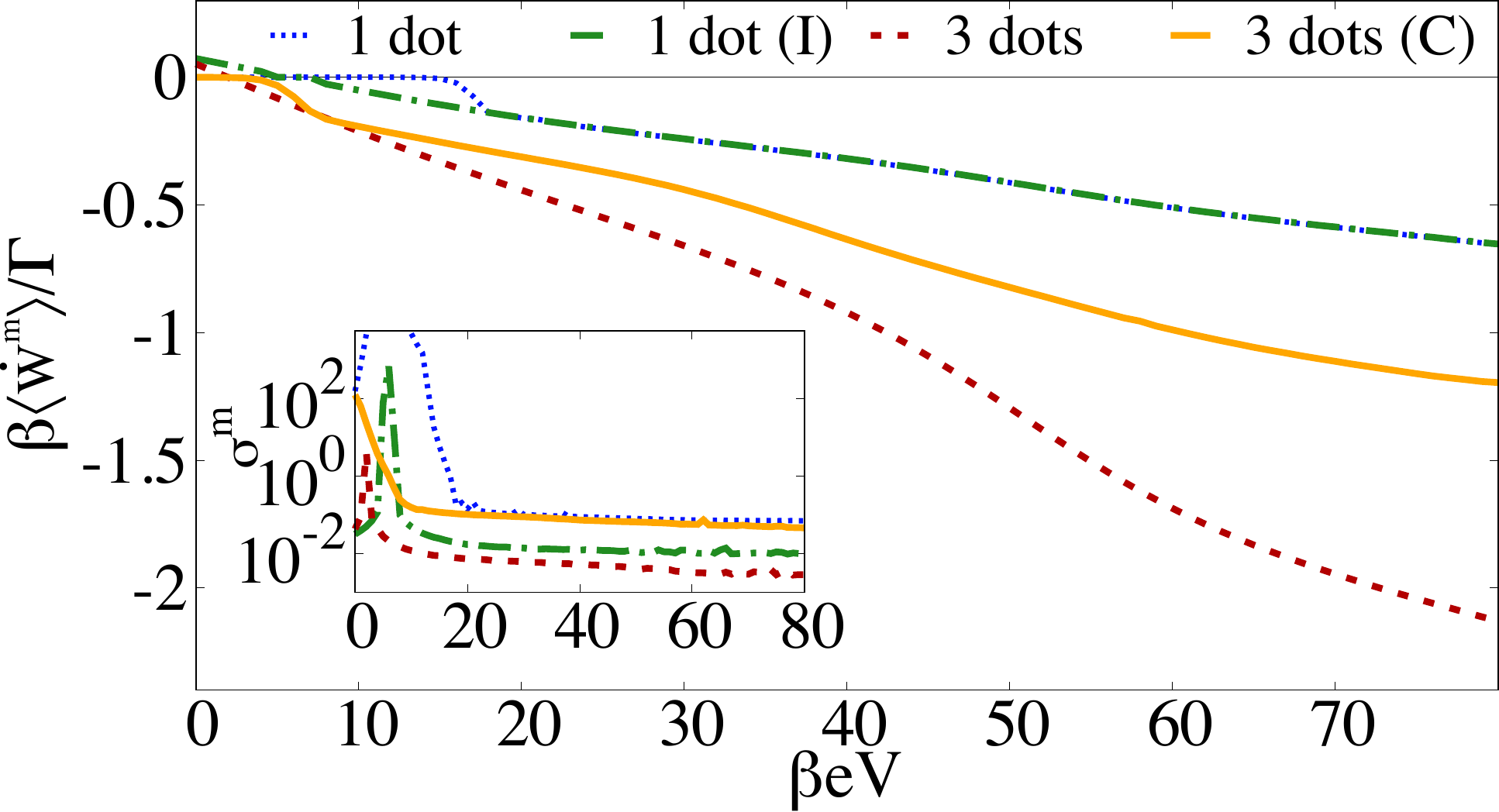}
\caption{Work output $\left<\dot W^{\text{m}}\right>$ as function of $\beta \text e V$ for $\beta M\text{g} r = 0.6$ for different engine setups (see main text). Inset: Normalized fluctuations $\sigma^{\text{m}}$.}
\label{fig:MultipleDots}
\end{figure}  

Lastly, we briefly discuss a possible up-scaling of the engine. This can be achieved by adding more QDs onto the rotor, optionally changing its length and therefore the moment of inertia $I$. We restrict the discussion to two 
extensions: First, adding two more QDs with rotational displacement $\Delta\phi = 2\pi/3$, which are transversally 
well separated by increasing the length of the rotor such that Coulomb and magnetic interactions between the moving QDs 
can be neglected. This setup is comparable to a multi-cylinder engine. Coulomb interactions can be neglected if the distance $d$ between two QDs is $d\gg 1\mu$m, hence, for a rotor of $l=30\mu$m and three equidistant QDs. Then, $I=9.6\cdot 10^{-36} \text{kg m}^2$. Second, adding two QDs onto the rotor at the original length and assuming that only one QD can be filled at a time due to strong Coulomb interaction. In Fig.~\ref{fig:MultipleDots} we plot $\left<\dot W^\text{m}\right>$ for $\beta M\text{g}r = 0.6$ as a function of $\beta \text{e}V$ for the first scenario (red dashed) and for the 
second (orange solid). For a better comparison we also plot the respective single QD versions for a rotor at the 
original length (blue dotted) and the increased length with larger $I$ (green dash-dotted). While the power output can be increased by roughly a factor $1.5$ (Coulomb, orange solid) or $3$ (no Coulomb, red dashed) as expected, 
$\sigma^{\text{m}}$ decreases by almost one order of magnitude (inset of Fig.~\ref{fig:MultipleDots}). Here, fluctuations decay due to the increased moment of inertia $I$ as well as the additional QDs. Note, that for the first proposal multiple dots can push the rotor simultaneously. We therefore assume that if one lifts the assumption of Coulomb blockade, i.e., allowing for multiple electrons on one QD, one will also increase the performance of our engine. Increasing the number of QDs further, which is only limited by the feasible length of the nanotube and possible friction effects due to additional bearings, and investigating the collective behaviour of the model therefore seems to be an interesting perspective for future work \cite{HerpichEtAlPRX2018, VroylandtEtAlEPL2018}.

\section{\label{sec:Conclusion}Conclusion}
Summarizing, we have provided a proposal and thermodynamical analysis of an autonomous stochastic engine. For experimentally realistic parameters we showed that the engine can transform chemical into mechanical work and vice versa with low fluctuations albeit moderate efficiency. Additionally, we clearly demonstrated the possibility to increase power and reduce fluctuation by increasing the number of QDs on the rotor. With the parameters used in this work, we can approximate the order of magnitude of the mass we can lift with the rotor engine: From Fig.~3 we find that $\beta M \text{g} r$ is in the order of magnitude of $1$, such that $M$ is about the order of magnitude of $10^{-15}$kg, which is about the mass of \emph{E. coli} \cite{NeidhardtBook1996}. 

\begin{acknowledgments}
The authors thank T. Brandes for initiating this project. C. W. acknowledges fruitful discussions with S. Restrepo. This work has been funded by the DFG - project 163436311 - SFB 910, the RTG 1558 and projects STR 1505/2-1 and BR 1528/8-2.
\end{acknowledgments}

\appendix 
\section{\label{secApp:DerivationTorque}Derivation of the torque acting on the QD}
In this section we derive the explicit forms of the torque $\tau_{\text{el}}(\phi,q) = -\alpha \text{e}V q r \sin(\phi)+\text{e}Uqr\cos(\phi)$ and $\tau_{\text{M}} = M\text{g}r$ acting on the rotor. The electric field generated by the two leads is taken to 
point in the $x$-direction (from left to right in Fig.~1), 
\begin{equation}
\mathbf E_{V} = \alpha \text{e}V \mathbf e_\text{x}, 
\end{equation}
where $\mathbf e_x$ is the unit vector along the $x$-axis. The externally applied electric field is pointing in the 
$y$-direction (from bottom to top in Fig.~1), such that it is given by
\begin{equation}
\mathbf E_U = \text{e}U\mathbf e_\text{y}
\end{equation}
with unit vector $\mathbf e_\text{y}$ in $y$-direction. The position $\mathbf x$ of the QD is given by 
$\mathbf x = r\cos(\phi)\mathbf e_\text{x} + r \sin(\phi)\mathbf e_\text{y}$. The electrostatic force $\mathbf F_{\text{el}} = q\mathbf E_\text{V} + q\mathbf E_\text{U}$ exerts a torque $\boldsymbol \tau_{\text{el}}$ on the QD, which is given by the cross product of $\mathbf x$ and the force, i.e., 
\begin{equation}
\begin{aligned}
\boldsymbol \tau_{\text{el}} &= \mathbf x \times \mathbf F_{\text{el}} = \left[\text{e}Uqr\cos(\phi)-\alpha\text{e} V q r\sin(\phi)\right]\mathbf e_\text{z} \\
&= \tau_{\text{el}}(\phi,q)\mathbf e_\text{z}.
\end{aligned}
\end{equation}
The force $\mathbf F_{\text{M}}$ stemming form the mass $M$ is always perpendicular to the position $\mathbf x$ of the QD, i.e., $\mathbf F_{\text{M}} = -M\text{g}\sin(\phi)\mathbf e_\text{x} + M\text{g}\cos(\phi)\mathbf e_\text{y}$, such that the torque exerted on the rotor is given by
\begin{equation}
\boldsymbol \tau_{\text{M}} = \mathbf x \times \mathbf F_{\text{M}} = M\text{g}r \mathbf e_\text{z} = \tau_{\text{M}}\mathbf e_\text{z}.
\end{equation}
Due to the fact, that both contributions to the total torque act along the $z$-axis, the dynamics can completely be described by an angle $\phi$ and angular velocity $\omega$. 

\section{\label{secApp:Trajectory}Trajectory representation}
Since the space of dynamical variables defined by the triple $(\phi,\omega,q)$ is large, we turn to the trajectory representation of Eq.~(1) in order to solve the steady state dynamics of the rotor engine numerically. The coupled stochastic differential equations 
\begin{align}
d\phi &= \omega dt,\label{eq:StochasticEquations1}\\
I d\omega &= \left[-\gamma \omega +\tau_{\text{el}}(\phi,q)+\tau_{\text{M}}\right]dt + \sqrt{2DI^2}dB(t),\label{eq:StochasticEquations2}\\
dq&= \sum_\nu dq^\nu =\sum\limits_{\nu q'}\left(q-q'\right)dN_{q'q}^\nu(\phi,t).\label{eq:StochasticEquations3}
\end{align}
reproduce the coupled Fokker-Planck and master Eq.~(1) at the ensemble level, which has been shown for a similar equation in \cite{WaechtlerEtAl}. In Eq.~(\ref{eq:StochasticEquations2}) $dB(t)$ denotes a Wiener process with mean $\mathbb E\left[dB(t)\right]=0$ and variance $\mathbb E\left[dB(t)^2\right] = dt$, which models the thermal fluctuations of the rotor. Here, $\mathbb E\left[\bullet\right]$ indicates an average over many realizations of the stochastic process. For a fixed $q$, Eqs.~(\ref{eq:StochasticEquations1}) and (\ref{eq:StochasticEquations2}) represent the Langevin equation for rotational Brownian motion of the rotor subject to the torque $\tau_{\text{el}}(\phi,q)$ and $\tau_{\text{M}}$. Eq.~(\ref{eq:StochasticEquations3}) describes the stochastic electron tunneling at the trajectory level, i.e., the change of the charge state $q$. The independent Poisson increments $dN_{q'q}^\nu(\phi,t)\in\{0,1\}$ obey the following statistics:
\begin{equation}
\label{eq:PoissonIncrement}
\begin{aligned}
\mathbb E\left[dN_{q'q}^\nu(\phi,t)\right] &= R^\nu_{q'q}(\phi) dt,\\
dN_{q'q}^\nu(\phi,t) dN_{\tilde qq}^{\tilde \nu}(\phi,t) &= \delta_{q'\tilde q}\delta_{\nu\tilde \nu} dN_{q'q}^{\nu}(\phi,t).
\end{aligned}
\end{equation}
The first equation shows that the average number of jumps into state $q'$ from a state $q$ in a time interval $dt$ is given by the tunneling rate $R^\nu_{q'q}(\phi)$. The second line of Eq.~(\ref{eq:PoissonIncrement}) enforces that at most one tunneling event per time interval can occur, i.e., either all $dN_{q'q}^\nu(\phi,t)$ are zero or $dN_{q'q}^\nu(\phi,t)=1$ for precisely one set of indices $q$, $q'$ and $\nu$. 

Additionally, we assume that the system is ergodic, such that we can sample the steady state probability density of the system by a single long trajectory. This also means that an ensemble average of an arbitrary quantity $A$ in the steady state is calculated by 
\begin{equation}
\left<A\right> = \frac{1}{T}\int\limits_0^T A(t)~dt,
\end{equation}
which is exact for ergodic systems in the limit of $T\to\infty$. 
We simulate the trajectories after a relaxation time of $\Gamma t=3000$ until $\Gamma T = 30,000,000$, where we have also checked that further relaxation time or simulation time does not change the probability density or averaged quantities. Note that we have also investigated different initial conditions and have not seen any dependency of the outcome on the initial conditions (after the relaxation time). Finally we note that the time step used in the simulations is $\Gamma \Delta t = 0.01$.

\section{\label{secApp:Thermodynamics}Laws of thermodynamics of the rotor engine}
In this section we derive the laws of thermodynamics of the engine. The total energy of the coupled system of rotor and QD is given by 
\begin{equation}
\label{eq:DefEnergy}
E = \frac{I\omega^2}{2} + \left[\varepsilon-\alpha\text{e} V r \cos(\phi)-\text{e}Ur\sin(\phi)\right]q,
\end{equation}
where the first term corresponds to the kinetic energy of the rotor and the second term to the energy of the QD in the electrostatic field generated by $V$ and $U$. The change in total energy of the system along a trajectory is either due to exchange of heat or to work performed on the system \cite{SekimotoBook2010}:
\begin{widetext}
\begin{equation}
\label{eqApp:firstLaw}
\begin{aligned}
dE =&I\omega \circ d\omega  + \left[\alpha \text{e}V r \sin(\phi) - \text{e}U r \cos(\phi)\right]\circ d\phi+ \left[\varepsilon-\alpha \text{e}V r \cos(\phi)-\text{e}Ur\sin(\phi)\right]\circ dq, 
\end{aligned}
\end{equation}
where $\circ$ denotes Stratonovich-type calculus. Furthermore, $dq = \sum_\nu dq^\nu$ and $dq^\nu=\sum_{q'}(q'-q)dN^\nu_{q'q}(\phi,t)$ denotes an electron jump with respect to reservoir $\nu$ [see Eq.~(\ref{eq:PoissonIncrement})]. By multiplying Eq.~(\ref{eq:StochasticEquations2}) with $\omega$, the second term of Eq.~(\ref{eqApp:firstLaw}) can be re-expressed, which yields
\begin{equation}
\label{eqApp:AfterFirstLaw}
\begin{aligned}
I\omega\circ d\omega &= \left[-\gamma \omega^2 -\alpha \text{e}V q r\sin(\phi)\omega + \text{e}Uqr \cos(\phi)\omega+M\text{g}r\omega\right]dt+ \sqrt{2DI^2}\omega\circ dB(t) \\
&=\left[-\alpha \text{e}V qr\sin(\phi)+\text{e}Uqr\cos(\phi)\right] \circ d\phi +M\text{g}r\omega dt-\gamma \omega^2 dt + \sqrt{2DI^2}\omega\circ dB(t).
\end{aligned}
\end{equation}
Inserting Eq.~(\ref{eqApp:AfterFirstLaw}) into Eq.~(\ref{eqApp:firstLaw}), we get
\begin{equation}
\label{eqApp:EnergyTrajectory}
\begin{aligned}
dE &= \left[\varepsilon - \alpha \text{e}V r\cos(\phi)-\text{e}Ur\sin(\phi)\right]\circ dq + M\text{g}r\omega dt- \gamma \omega^2 dt + \sqrt{2DI^2 }\omega\circ dB_t. \\
\end{aligned}
\end{equation}
\end{widetext}
As the first law states that changes in the total energy are due to heat exchange or due to work performed on the system, i.e., 
\begin{equation}
dE = \sum\limits_\nu \delta Q^\nu+ \delta Q^\text{rot} +\delta W^\text{chem} + \delta W^{\text{mech}},
\end{equation}
we identify the different contributions as follows: The chemical work $\delta W^\text{chem}=\sum_\nu \mu^\nu dq^\nu$ and the mechanical work $\delta W^{\text{mech}}=M\text{g}r\omega dt$. The heat flow to the rotor from its thermal reservoir due to friction and thermal noise is given by $\delta Q^\text{rot}=-\gamma \omega^2dt+\sqrt{2DI^2}\omega\circ dB(t)$ \cite{SekimotoBook2010}. 
The remaining terms in Eq.~(\ref{eqApp:EnergyTrajectory}) are identified as heat exchanged with the reservoir $\nu$, defined as $\delta Q^\nu = \left[\varepsilon- \alpha \text{e}V r\cos(\phi)-\text{e}Ur\sin(\phi)-\mu^\nu\right]\circ dq^\nu$. With these definitions of heat and work we can derive a consistent second law as we will show later in this section.

By averaging of Eq.~(\ref{eqApp:EnergyTrajectory}) over many realizations, equivalently averaging with respect to the probability density \cite{WaechtlerEtAl}, we find
\begin{equation}
\label{eqApp:FirstLawEnsemble}
\left<\frac{dE}{dt}\right> = \sum\limits_\nu \left<\dot Q^\nu\right> + \left<\dot Q^\text{rot}\right> + \left<\dot W^\text{chem}\right> + \left<\dot W^{\text{mech}}\right>.
\end{equation}
Here, the averaged heat exchange with reservoir $\nu$ takes the form
\begin{equation}
\label{eq:DefHeat}
\left<\dot Q^\nu\right> = \left(\varepsilon - \mu^\nu\right)\left<I_{\text{M}}^\nu\right> -\alpha \text{e}V r\left<\cos(\phi)I_{\text{M}}^\nu\right> - \text{e}Ur\left<\sin(\phi)I_{\text{M}}^\nu\right>.
\end{equation}
The average matter current from reservoir $\nu$, $\left<I_{\text{M}}^\nu\right> \equiv \mathbb E\left[dq^\nu/dt\right]$, 
can be equivalently expressed via the ensemble average 
\begin{equation}
\label{eq:I_avg}
\left<I_{\text{M}}^\nu\right> = \int d\phi d\omega \left[ R^\nu_{10}(\phi)p_0-R^\nu_{01}(\phi)p_1\right] ,
\end{equation}
as well as correlations of $\sin(\phi)$ and the current,
\begin{equation}
\label{eq:sinI_avg}
\left<\sin(\phi)I_{\text{M}}^\nu\right> = \int d\phi d\omega~\sin(\phi)\left[R^\nu_{10}(\phi)p_0-R^\nu_{01}(\pi)p_1\right].
\end{equation}
The correlations of $\cos(\phi)$ and the current are defined equivalently by replacing $\sin$ with $\cos$ in Eq.~(\ref{eq:sinI_avg}). 
The heat current entering from the reservoir of the rotor is given by
\begin{equation}
\label{eq:DefinitonHeatOsc}
\left<\dot Q^\text{rot}\right> = -\gamma \left(\left<\omega^2\right> - \frac{1}{I\beta}\right).
\end{equation}
and the averaged chemical work and mechanical work corresponding to lifting or lowering the weight $M$ reads as in Eq.~(4).
Note, that the mechanical average power $\left<\dot W^{\text{mech}}\right>$ is directly proportional to the average velocity of the rotor, i.e., if $\left<\omega\right> >0$ mechanical work is performed on the system and if $\left<\omega\right><0$ the engine performs work by lifting the weight.

To establish that the second law holds, i.e., that the average total entropy production rate is non-negative, we consider the evolution of the Shannon entropy 
\begin{equation}
\label{eq:S(t)}
S(t)= -\int d\phi d\omega\sum\limits_{q}p_q(\phi,\omega;t) \ln p_q(\phi,\omega;t), 
\end{equation}
where $p_q(\phi,\omega;t)$ is the solution of the coupled Fokker-Planck and master Eq.~(1). Taking the time derivative of $S(t)$, introducing the shorthand notation $p_q\equiv p_q(\phi,\omega;t)$, and $\sumint \equiv \int d\phi \int d\omega \sum_q$, and using the conservation of probability as well as partial integration (assuming vanishing boundary contributions, $\lim_{\phi\to\pm\infty} \phi p = \lim_{\omega\to\pm\infty} \omega p = 0$) we obtain
\begin{equation}
\label{eq:StochasticEntropy}
\begin{aligned}
\frac{d}{dt}S(t) =& \underbrace{\sumint \left[\partial_\omega J_q(\phi,\omega,t)\right] \ln p_q}_{\equiv \dot S_1(t)} \underbrace{- \sumint \sum\limits_{q'\nu}R_{qq'}^\nu(\phi) p_{q'} \ln p_q}_{\equiv \dot S_2(t)},
\end{aligned}
\end{equation}
where 
\begin{equation}
J_q (\phi,\omega,t) = -\frac{\gamma}{I} \omega p_q - D\partial_\omega p_q
\end{equation}
is a probability current. 
We integrate by parts to rewrite $\dot S_1(t)$ as follows:
\begin{equation}
\label{eq:S1dot}
\dot S_1(t) =  \sumint \left[ \frac{\gamma}{I} \omega \partial_\omega p_q + D \frac{(\partial_\omega p_q)^2}{p_q} \right] .
\end{equation}
From Eq.~(\ref{eq:DefinitonHeatOsc}) we obtain
\begin{equation}
\label{eq:0}
0 = \beta \left<\dot Q^\text{rot}\right> + \sumint \left( \beta \gamma \omega^2 p_q + \frac{\gamma}{I} \omega\partial_\omega p_q \right) .
\end{equation}
Summing Eqs.~(\ref{eq:S1dot}) and (\ref{eq:0}) 
we arrive at
\begin{equation}
\label{eq:S1dot_final}
\dot S_1(t) = \beta \left<\dot Q^\text{rot}\right> + \dot \Sigma_{\text{cont}} ,
\end{equation}
where
\begin{equation}
\label{eq:EntropyProductionContinuous}
 \dot \Sigma_{\text{cont}}  =  \sumint \frac{\left[\gamma \omega p_q + DI\partial_\omega p_q\right]^2}{DI^2p_q} \geq 0.
\end{equation}

Next, we rewrite the second term on the right side of Eq.~(\ref{eq:StochasticEntropy}) as follows:
\begin{equation}
\label{eq:S2dot}
\dot S_2(t) = -\frac{1}{2} \sumint \sum\limits_{q'\nu}\left(R_{qq'}^\nu p_{q'} \ln p_q + R_{q'q}^\nu p_q\ln p_{q'}\right) .
\end{equation}
From the property of detailed balance obeyed by the electron tunneling rates [see Eq.~(\ref{eq:Rates})], i.e.
\begin{equation}
\label{eq:LocalDetailedBalance}
\frac{R_{01}^\nu}{R_{10}^\nu} = e^{\beta\left[\varepsilon - \alpha\text{e} V r \cos(\phi)-\text{e}Ur\sin(\phi) -\mu^\nu\right]},
\end{equation}
we derive the identity
\begin{equation}
\label{eq:identity}
0 = 
\beta\sum_\nu  \left<\dot Q^\nu\right>  - \frac{1}{2} \sumint \sum\limits_{q'\nu} \left( R_{qq'}^\nu p_{q'} - R_{q'q}^\nu p_q \right) \ln\frac{R_{q'q}^\nu}{R_{qq'}^\nu} , \\
\end{equation}
where the first term on the right relates to heat exchange with the fermionic leads [see Eqs.~(\ref{eq:DefHeat}) - (\ref{eq:sinI_avg})].
Summing Eqs.~(\ref{eq:S2dot}) and (\ref{eq:identity}) and rearranging terms, we obtain
\begin{equation}
\label{eq:S2dot_final}
\dot S_2(t) = \beta \sum\limits_\nu \left<\dot Q^\nu\right> + \dot \Sigma_{\text{disc}},
\end{equation}
where
\begin{equation}
\label{eq:EntropyProductionDiscrete}
\dot \Sigma_{\text{disc}} = \frac{1}{2} \sumint \sum\limits_{q'\nu}\left(R_{qq'}^\nu p_{q'} - R_{q'q}^\nu p_q\right)\ln\frac{R_{qq'}^\nu p_{q'}}{R_{q'q}^\nu p_q}  \geq 0 .
\end{equation}
Here, non-negativity follows from the log-sum inequality. Adding Eqs.~(\ref{eq:S1dot_final}) and (\ref{eq:S2dot_final}), we find that the total entropy production rate is given by
\begin{equation}
\label{eq:EntropyAll}
\dot \Sigma = \dot \Sigma_{\text{cont}} + \dot \Sigma_{\text{disc}} = \frac{d}{dt}  S - \beta\left(\left<\dot Q^\text{rot}\right> + \sum\limits_\nu \left<\dot Q^\nu\right>\right) \geq 0, 
\end{equation}
where the non-negativity of $\dot\Sigma$ shows, that the second law holds in our system.

At steady state, $\partial S/\partial t=0$ and $\left<dE/dt\right>=0$, the second law, Eq.~(\ref{eq:EntropyAll}), becomes Eq.~(5).

\end{document}